\documentclass[aps,prb,floatfix,twocolumn,showpacs,preprintnumbers,superscriptaddress,amsmath,amssymb]{revtex4-1}

\usepackage{graphicx}  
\usepackage{dcolumn}   
\usepackage{bm}        
\usepackage{color}

\begin{document}

\preprint{}

\title{Magnetoplasmonic properties of perpendicularly magnetized $[\textnormal{Co}/\textnormal{Pt}]_{N}$ nanodots}

\author{Francisco Freire-Fern{\'a}ndez} \email{francisco.freirefernandez@aalto.fi}
\affiliation{NanoSpin, Department of Applied Physics, Aalto University School of Science, P.O. Box 15100, FI-00076 Aalto, Finland}
\author{Rhodri Mansell}
\affiliation{NanoSpin, Department of Applied Physics, Aalto University School of Science, P.O. Box 15100, FI-00076 Aalto, Finland}
\author{Sebastiaan van Dijken} \email{sebastiaan.van.dijken@aalto.fi}
\affiliation{NanoSpin, Department of Applied Physics, Aalto University School of Science, P.O. Box 15100, FI-00076 Aalto, Finland}

\date{\today}
	
\begin{abstract}
We demonstrate a ten-fold resonant enhancement of magneto-optical effects in perpendicularly magnetized $[\textnormal{Co}/\textnormal{Pt}]_{N}$ nanodots mediated by the excitation of optimized plasmon modes. Two magnetoplasmonic systems are considered; square arrays of $[\textnormal{Co}/\textnormal{Pt}]_{N}$ nanodots on glass and identical arrays on a Au/Si$\textnormal{O}_{2}$ bilayer. On glass, the optical and magneto-optical spectra of the nanodot arrays are dominated by the excitation of a surface lattice resonance (SLR), whereas on Au/Si$\textnormal{O}_{2}$, a narrow surface plasmon polariton (SPP) resonance tailors the spectra further. Both the SLR and SPP modes are magneto-optically active leading to an enhancement of the Kerr angle. We detail the dependence of optical and magneto-optical spectra on the number of Co/Pt bilayer repetitions, the nanodot diameter, and the array period, offering design rules on how to maximize and spectrally tune the magneto-optical response of perpendicularly magnetized $[\textnormal{Co}/\textnormal{Pt}]_{N}$ nanodots. 
\end{abstract}
	
\maketitle

\section{Introduction}
\label{Sec:Intro}

Perpendicularly magnetized Co/Pt bilayers and multilayers are widely investigated in the fields of nanomagnetism and spintronics. While research was motivated initially by their potential use in magnetic data storage devices\cite{JOH-96,TER-05}, Co/Pt and other similar structures have been instrumental also in studies on domain wall dynamics\cite{LEM-98,MET-07}, current-induced magnetization switching\cite{MAN-06,MIR-11,LIU-12}, current-driven motion of chiral domain walls\cite{HAA-13,EMO-13,RYU-13}, ionic control of magnetism\cite{BAU-15}, and magnetic skyrmions\cite{WOO-16,BOU-16,MOR-16}. The attractive properties of Co/Pt arise from the interface nature of its perpendicular magnetic anisotropy (PMA)\cite{NAK-98,HEL-17}, allowing it to be tailored by variation of the Co layer thickness or interface chemistry and, in combination with the Dzyaloshinskii-Moriya interaction (DMI), it can be used to stabilize chiral spin textures. Moreover, the demonstration of all-optical switching of perpendicular magnetization in Co/Pt multilayers\cite{LAM-14} has triggered a lively debate on the origin of AOS in thin ferromagnetic films.

The magneto-optical properties of Co/Pt multilayers are used mainly for magnetic characterization, often in the polar Kerr effect configuration\cite{ZEP-89}. The complex refractive index and magneto-optic Voigt parameter determine the optical and magneto-optical response of continuous Co/Pt films. The gradual variation of both parameters with wavelength produces a rather smooth magneto-optical Kerr effect (MOKE) spectrum\cite{ATK-96}. In patterned nanostructures, this no longer necessarily holds true. If Co/Pt multilayers are patterned into dots wherein the resonance condition of the free electrons matches the wavelength of incident light, a localized surface plasmon resonance (LSPR) is excited. For other magnetic nanostructures, it has been demonstrated already that the optical near-fields of LSPR modes can resonantly enhance and change the polarity of magneto-optical signals\cite{CHE-11,BON-11,ARM-13}. This phenomenon is explained by the excitation of two electric dipoles (LSPRs) within the nanodots. The first dipole is excited parallel to the incident electric field. The second dipole is excited orthogonally to both the first dipole and the direction of the magnetization of the nanodot, and is induced by spin-orbit coupling\cite{MAC-13}. Because the phase and amplitude relations of the two electric dipoles determine the magneto-optical response, it no longer depends solely on intrinsic material parameters. Tailoring of optical near-fields by variation of the nanodot size, shape, or their ordering in periodic arrays, as routinely exploited in plasmonics, can therefore be used to design magneto-optical spectra\cite{MAC-13,KAT-15,MAC-16}.    

Compared to metallic plasmonic systems comprising Ag, Au, or Al, magnetic metals suffer from larger ohmic losses because of their higher electrical resistivity. Consequently, plasmonic resonances of magnetic nanostructures are broader and less intense. To mitigate losses, hybrid magnetoplasmonic materials combining noble and magnetic metals have been explored. Examples include, Au/Co/Au trilayers\cite{TEM-10}, nanosandwiches\cite{GON-08}, or nanorods\cite{ARM-16}, and core-shell Co/Ag or Co/Au nanoparticles\cite{WAN-11,SON-12}. The Co/Pt multilayers, considered in this paper would not, a priori, circumvent losses in a similar fashion because the electrical resistivity of Pt is higher than that of Co. Consequently, it remains an open question to what extent plasmon resonances could be exploited to tailor the magneto-optical response of this attractive PMA system. The ability to drastically enhance optical near-fields in Co/Pt nanostructures using plasmonics would provide promising links to ongoing research on spintronics and AOS of magnetic materials. 

\begin{figure*}
	\includegraphics{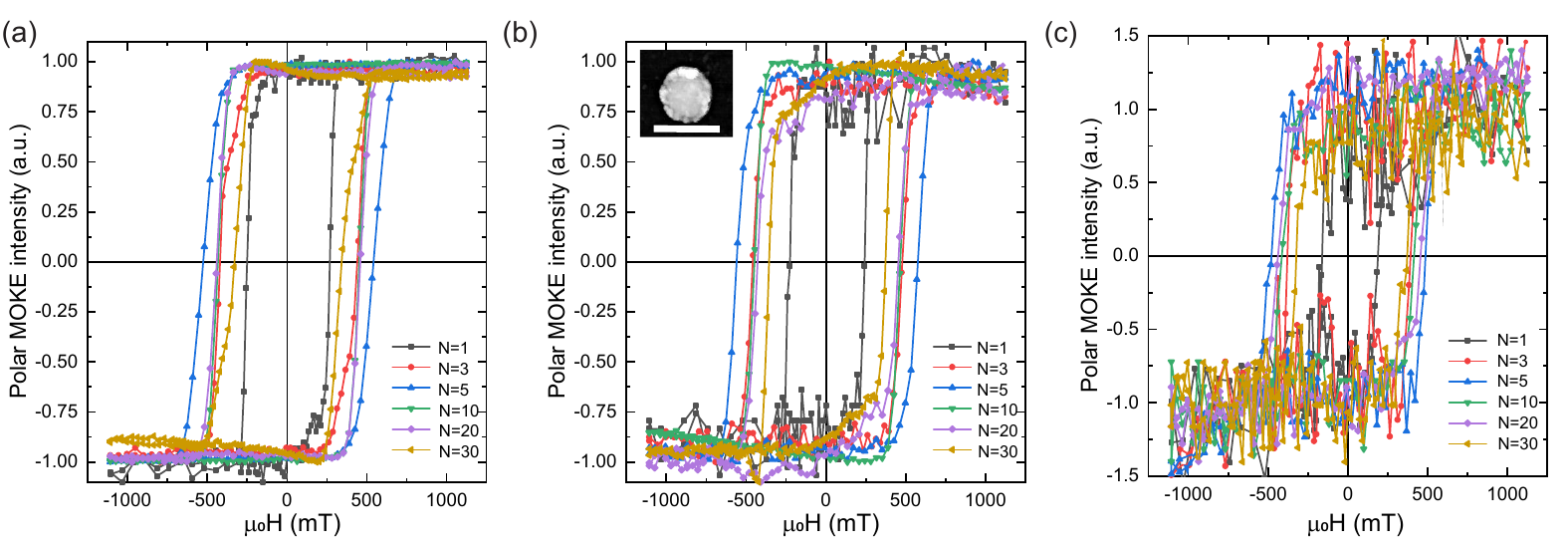}
	\caption{Polar MOKE hysteresis curves of square arrays of $[\textnormal{Co}/\textnormal{Pt}]_{N}$ nanodots with a diameter of (a) 200 nm, (b) 150 nm, and (c) 100 nm. The array period ($P$) is 400 nm and the number of bilayers repetitions ($N$) varies from 1 to 30. The inset in panel (b) shows an atomic force microscopy image of a $[\textnormal{Co}/\textnormal{Pt}]_{10}$ nanodot with a diameter of 150 nm. The white scale bar corresponds to 200 nm.}
	\label{Fig1}
\end{figure*}

Here, we study the optical and magneto-optical properties of [Co/Pt]$_N$ nanodots. We consider two magnetoplasmonic systems; (i) periodic [Co/Pt]$_N$ nanodot arrays on glass substrates and (ii) identical nanodot arrays on Au/Si$\textnormal{O}_{2}$ bilayers. We demonstrate that both plasmonic nanostructures allow for the design of strong magneto-optical responses through the excitation of collective surface lattice resonances (SLRs, system (i) and (ii)) or the excitation of surface plasmon polaritons (SPPs, only system (ii)). Design rules for strong magnetoplasmonic effects in this PMA system are derived by characterizing a large number of samples with varying numbers of bilayer repetitions ($N$), nanodot diameters ($D$), and array periods ($P$).                    

\section{Experiment}
\label{Sec:Exp}

The periodic arrays in this study consist of $\textnormal{Ta}(2)/\textnormal{Pt}(4)/[\textnormal{Co}(1)/\textnormal{Pt}(1)]_{N}/\textnormal{Pt}(2)$ multilayer nanodots, where the numbers in brackets indicate the layer thickness in nanometers. Hereafter, we will simply refer to the multilayer structure as $[\textnormal{Co}/\textnormal{Pt}]_{N}$. The nanodot arrays are fabricated on glass substrates and Si substrates covered by a Au(150)/Si$\textnormal{O}_{2}(20)$ bilayer using electron beam lithography and lift-off. The $[\textnormal{Co}/\textnormal{Pt}]_{N}$ multilayers are grown by magnetron sputtering, whereas Si$\textnormal{O}_{2}$ and Au are deposited by atomic layer deposition and electron beam evaporation, respectively. Using this nanofabrication process, square arrays of $[\textnormal{Co}/\textnormal{Pt}]_{N}$ nanodots with the following parameters are patterned on glass and Si/Au/Si$\textnormal{O}_{2}$: The number of bilayer repetitions is varied as $N$ = {1, 3, 5, 10, 20, 30}. For each $N$, arrays with periods $P$ = {350 nm, 400 nm, 450 nm, 500 nm} are patterned. Finally, samples with nanodot diameters $D$ = 100 nm, 150 nm, and 200 nm are fabricated for each combination of $N$ and $P$.   

The optical and magneto-optical properties of the samples are characterized in a magneto-optical spectrometer that can be configured for transmission (Faraday effect) and reflection (Kerr effect) measurements. The setup consists of a NKT SuperK EXW-12 supercontinuum laser with an acousto-optical filter, polarizing and focusing optics, a Hinds Instruments I/FS50 photoelastic modulator, and a photodetector. The laser wavelength is varied between 475 nm and 1050 nm. We use linear polarized light at normal incidence with the electric field aligned along one the $x$-axis of the square nanodot arrays. During measurements, a $\pm$1 T field from an electromagnet switches the magnetization of the $[\textnormal{Co}/\textnormal{Pt}]_{N}$ nanodots between two perpendicular directions. The magneto-optical Kerr rotation ($\theta$) and Kerr ellipticity ($\epsilon$) are simultaneously recorded by lock-in amplification of the modulated signal at 50 kHz and 100 kHz. From these data the Kerr angle ($\Phi$) is calculated as $\Phi=\sqrt{{\theta}^2+{\epsilon}^2}$. All measurements are performed with the $[\textnormal{Co}/\textnormal{Pt}]_{N}$ nanodots immersed in oil. The refractive index of the oil matches that of the glass substrate ($n$ = 1.52). The resulting uniform dielectric environment facilitates the excitation of intense plasmon resonances.   

\begin{figure*}
	\includegraphics{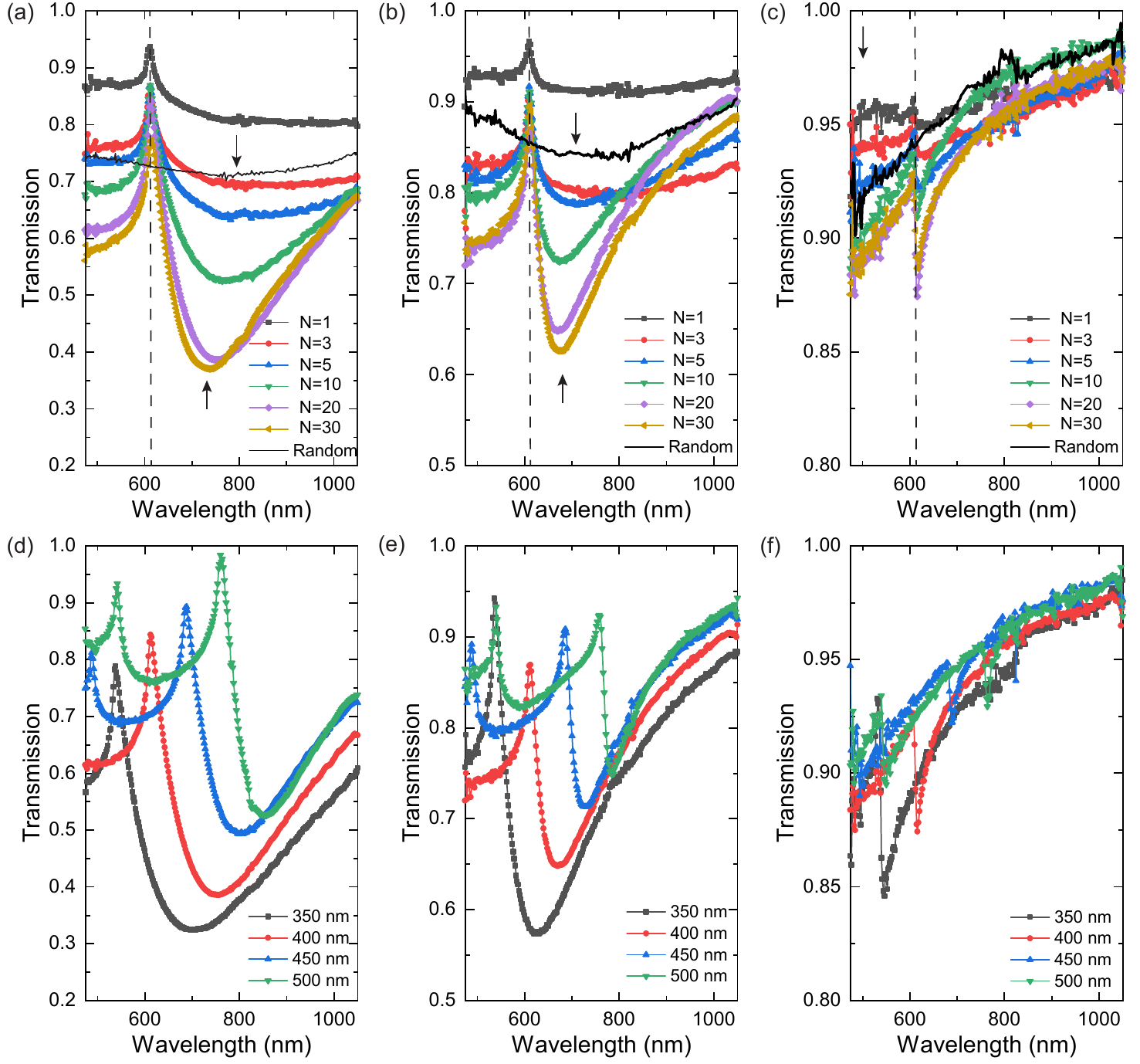}
	\caption{Optical transmission spectra of square arrays of $[\textnormal{Co}/\textnormal{Pt}]_{N}$ nanodots as a function of $N$ (a-c) and $P$ (d-f). Panels (a-c) show data for arrays with $P$ = 400 nm and $D$ = 200 nm (a), 150 nm (b), and 100 nm (c). The black lines represent spectra of randomly distributed $[\textnormal{Co}/\textnormal{Pt}]_{20}$ nanodots with the same packing density. The vertical dashed lines denote the position of the (0,$\pm$1) DOs, the downward pointing arrows indicate the single-particle LSPRs, and the upward pointing arrows mark the SLR wavelengths. Panels (d-f) depict data for arrays with $N$ = 20 nm and $D$ = 200 nm (d), 150 nm (e), and 100 nm (f).}
	\label{Fig2}
\end{figure*}

\section{Results and discussion}
\label{Sec:Results} 
\subsection{Magnetic characterization}  

The magnetic properties of the nanodots are strongly influenced by the number of Co/Pt bilayer repetitions. Figure \ref{Fig1} shows polar MOKE hysteresis curves as a function of $N$ and $D$. As is expected, the Co/Pt multilayers display an out-of-plane magnetic easy axis due to hybridization of the electron orbitals at the Co/Pt interfaces\cite{NAK-98,HEL-17}. The hysteresis curves in Figs. \ref{Fig1}(a)-(c) exhibit fully remanent magnetization for most values of $N$, with only a slight decrease for the thickest multilayers. Regardless of the nanodot diameter, the coercivity of the arrays displays a similar trend as a function of $N$. The coercivity increases initially and reaches a maximum for five bilayer repetitions. Beyond this, it slowly decreases again in the thicker nanodots. The Co layers are ferromagnetically exchanged coupled to each other through the 1 nm thick Pt layers\cite{KNE-05}, which leads to increasing coercivity with increasing $N$. For thicker multilayers, larger dipolar fields allow nucleation of reverse domains at lower fields, leading to a reduction in coercivity and remanence\cite{SBI-10}.

\subsection{Optical and magneto-optical properties of $[\textnormal{Co}/\textnormal{Pt}]_{N}$ nanodot arrays on glass}  

Firstly, we discuss the optical response of $[\textnormal{Co}/\textnormal{Pt}]_{N}$ nanodot arrays on glass. Figures \ref{Fig2}(a)-(c) show optical transmission spectra of $[\textnormal{Co}/\textnormal{Pt}]_{N}$ arrays with $P$ = 400 nm and $D$ = 200 nm (a), 150 nm (b), and 100 nm (c). Nanodots of these sizes support the excitation of LSPRs at visible wavelengths, as the solid lines recorded on randomly distributed $[\textnormal{Co}/\textnormal{Pt}]_{20}$ nanodots demonstrate. The downward pointing arrows mark the LSPR wavelengths. The LSPRs are broad because of large ohmic losses in Co and Pt. In periodic nanodot arrays, coupling between the single-particle LSPRs and the diffracted orders (DOs) of the array produces asymmetric Fano-like excitations, known as surface lattice resonances (SLRs)\cite{ZOU-04,KRA-08,AUG-08,KRA-18}. In uniform dielectric environments (glass/index matching oil in our experiment, $n$ = 1.52), the DO wavelengths are given by

\begin{equation}
\lambda_{p,q}=\frac{nP}{\sqrt{{p^2+q^2}}},
\label{Eq1}
\end{equation}

\noindent where $p$ and $q$ indicate the order of diffraction along the $x$- and $y$-axis of the nanodot array. In optical transmission spectra, the DOs appear as sharp peaks. For instance, the vertical dashed lines in Figs. \ref{Fig2}(a)-(c) mark the (0,$\pm$1) DOs of arrays with $P$ = 400 nm ($\lambda_{0,\pm{1}}=1.52\times400=608$ nm). 

Collective SLR modes absorb the incident light efficiently and, hence, they produce a minimum in optical transmission spectra. The upward pointing arrows in Figs. \ref{Fig2}(a) and \ref{Fig2}(b) indicate the SLRs arising from hybridization between the single-particle LSPRs and the (0,$\pm$1) DOs of the array. Because larger and thicker nanodots absorb more light, the transmission signal at the SLR wavelength decreases with increasing $D$ and $N$. The SLRs are more narrow than the LSPRs because diffraction in the array plane produces scattered fields that counter damping of the single-particle plasmonic response\cite{KRA-18}. In our $[\textnormal{Co}/\textnormal{Pt}]_{N}$ nanodot arrays, the SLR linewidth decreases as the number of bilayer repetitions increases. Moreover, changes in the size and aspect ratio of the nanodots translate into a spectral shift of the LSPR\cite{BAR-16} and, thereby, the SLR mode. For instance, Figs. \ref{Fig2}(a) and \ref{Fig2}(b) show how an increase of $N$ and a decrease of $D$ blue-shift the SLR transmission minimum. For $[\textnormal{Co}/\textnormal{Pt}]_{N}$ nanodots with $D$ = 100 nm, the LSPR is blue-shifted well below the DO wavelength. Together with the reduced polarizability of this small-volume particle, it prevents the excitation of a pronounced SLR mode. 

\begin{figure*}
	\includegraphics{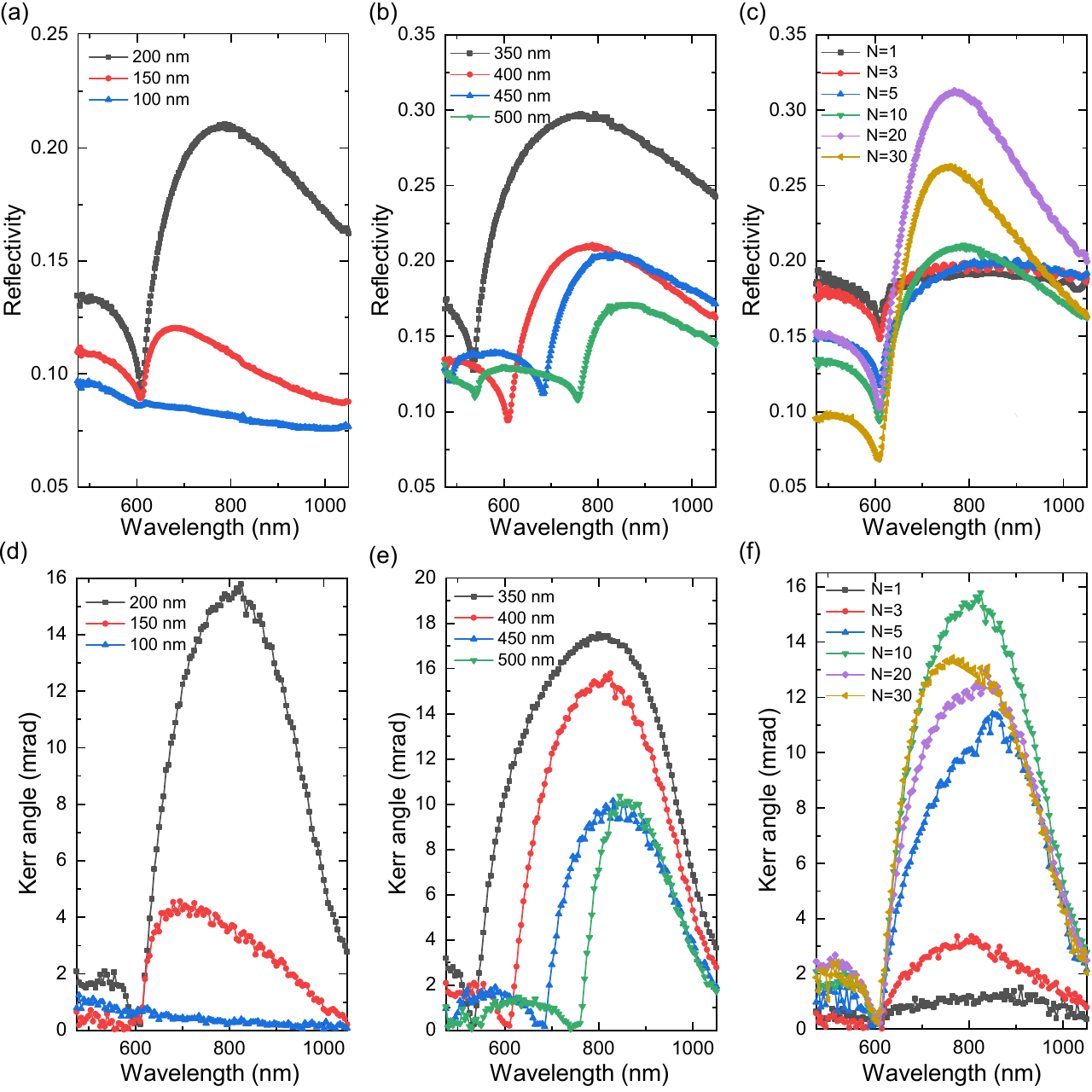}
	\caption{Optical reflectivity spectra (a-c) and polar MOKE spectra (d-f) of square arrays of $[\textnormal{Co}/\textnormal{Pt}]_{N}$ nanodots as a function of $D$ (a,d), $P$ (b,e), and $N$ (c,f). In (a,d) $N$ = 10 and $P$ = 400 nm, in (b,e) $N$ = 10 and $D$ = 200 nm, and in (c,f) $P$ = 400 nm and $D$ = 200 nm.}
	\label{Fig3}
\end{figure*}

Figures \ref{Fig2}(d)-(f) summarize the dependence of optical transmission spectra on the array period. An increase of $P$ red-shifts the DOs, as described by Eq. \ref{Eq1}. Consequently, the SLR wavelength moves up too. For nanodot arrays with $P$ = 450 nm and 500 nm, sharp transmission peaks at the spectral position of ($\pm$1,$\pm$1) DOs are visible below 600 nm also. The SLR linewidth depends on the spectral overlap between the LSPR and DO modes\cite{ZOU-04,KRA-08,AUG-08,KRA-18}. In our $[\textnormal{Co}/\textnormal{Pt}]_{20}$ samples with $D$ = 200 nm and 150 nm, a red-shift of the DO towards the single-particle LSPR reduces the SLR linewidth. The results of Fig. \ref{Fig2} demonstrate that $[\textnormal{Co}/\textnormal{Pt}]_{N}$ nanodot arrays support the excitation of intense SLRs if $D\geq$ 150 nm and $N\geq$ 5. The minimal SLR linewidth is about 100 nm, which is comparable to that of previously studied magnetoplasmonic arrays made of Ni nanodisks\cite{KAT-15,MAC-16,POU-18} or Au/Si$\textnormal{O}_{2}$/Ni dimers\cite{POU-18}.  

The optical reflectivity spectra of Figs. \ref{Fig3}(a)-(c) validate the excitation of SLRs in the $[\textnormal{Co}/\textnormal{Pt}]_{N}$ nanodot arrays. In this measurement geometry, scattering from the SLR mode into the far field enhances the reflectivity ($R$) at the SLR wavelength. We note that the optical reflectivity of a nanodot array does not exactly correspond to 1 - $T$, because $R$ depends on scattering by the nanodots, while the optical transmission $T$ is affected by the scattering and absorption of light. SLRs do not only determine the optical response of the $[\textnormal{Co}(1)/\textnormal{Pt}(1)]_{N}$ nanodot arrays, but also their magneto-optical activity. In our experiments, we irradiate the nanodots with linear polarized light at normal incidence. In this geometry, the incident electric field ($E_\mathrm{x}$) induces an oscillating electric dipole ($p_\mathrm{x}$) in the metal nanodots, corresponding to the LSPR mode. The electric dipole is given by $p_\mathrm{x}=\alpha_\mathrm{xx}E_\mathrm{x}$, where $\alpha_\mathrm{xx}$ is a diagonal component of the nanodot polarizability tensor. In the presence of perpendicular magnetization, spin-orbit coupling produces a second weaker electric dipole orthogonal to the optically excited dipole and the direction of magnetization ($p_\mathrm{y}=\alpha_\mathrm{xy}E_\mathrm{x}$). For single or randomly distributed magnetic nanodots, the real and imaginary part of the $p_\mathrm{y}/p_\mathrm{x}$ ratio determine to Kerr rotation and Kerr ellipticity, respectively\cite{MAC-13}. 

The optical reflectivity of a periodic plasmonic array is proportional to the effective polarizability squared ($R\propto|\alpha_\mathrm{eff}|^2$)\cite{BOH-83}. The effective polarizability of nanodots in an array accounts for polarizing effects caused by the incident radiation and the electric fields from other nanodots in the array. For normal incident light with linear polarization along the $x$-axis ($E_\mathrm{x}$), the relevant diagonal component of the effective polarizability tensor ($\alpha_\mathrm{eff,xx}$) can be written as\cite{AUG-08,KRA-18}

\begin{equation}
\alpha_\mathrm{eff,xx} = \frac{1}{1/\alpha_\mathrm{xx} - S_\mathrm{x}}
\label{Eq2}
\end{equation}      

\noindent In Eq. \ref{Eq2}, $\alpha_\mathrm{xx}$ is the polarizability of a single metal nanodot and $S_\mathrm{x}$ is the geometrical lattice factor for an incoming electric field along $x$. The effective polarizability of a plasmonic array is thus resonantly enhanced when the real part of the denominator in Eq. \ref{Eq2} (Re($1/\alpha_\mathrm{xx}$) - Re($S_\mathrm{x}$), becomes zero. This condition corresponds to the SLR wavelength. The SLR linewidth depends on the imaginary value of the denominator in Eq. \ref{Eq2}. As the polarizability of magnetic nanodots is small compared to that of noble metals, i.e. $1/\alpha_\mathrm{xx}$ is large, the SLRs of magnetoplasmonic arrays tend to be substantially broader than their noble metal counterpart. In magnetic nanodot arrays with perpendicular magnetization, $E_\mathrm{x}$ induces a second orthogonal SLR mode through spin-orbit coupling. This mode can be thought of as arising from diffractive coupling of the local electric dipoles along $y$ ($p_\mathrm{y}$). The effective polarizability of the orthogonal SLR mode is given by\cite{MAC-16,POU-19}

\begin{equation}
\alpha_\mathrm{eff,xy} = \frac{\alpha_\mathrm{xy}}{\alpha_\mathrm{xx}\alpha_\mathrm{yy}(1/\alpha_\mathrm{yy} - S_\mathrm{x})(1/\alpha_\mathrm{xx} - S_\mathrm{y})}, 
\label{Eq3}
\end{equation}

\noindent where $\alpha_\mathrm{eff,yy}$ is the second diagonal component of the effective polarizability tensor of the magnetic nanodot array, $\alpha_\mathrm{eff,xy}$ is the off-diagonal component, and $S_\mathrm{y}$ is the lattice factor for radiation along $y$. The magneto-optical Kerr angle of the array ($\Phi$) is then given by 

\begin{equation}
\Phi = \Bigg|\frac{\alpha_\mathrm{eff,xy}}{\alpha_\mathrm{eff,xx}}\Bigg|. 
\label{Eq4}
\end{equation}

For square arrays of circular magnetic nanodots,  $\alpha_\mathrm{xx}=\alpha_\mathrm{yy}$ and $S_\mathrm{x}=S_\mathrm{y}$. From Eqs. \ref{Eq2} - \ref{Eq4} it follows that the spectral positions of the DOs and SLRs in optical reflectivity and MOKE spectra are similar in this case. At the SLR wavelength, the Kerr rotation is maximized and the Kerr ellipticity crosses zero\cite{KAT-15}. As a result, the maximum Kerr angle ($\Phi=\sqrt{{\theta}^2+{\epsilon}^2}$) may be slightly shifted away from the SLR wavelength. If the symmetry of the nanodots or the array is broken, the reflectivity and MOKE spectra can be vastly different, as demonstrated experimentally for circular Ni nanoparticles in rectangular arrays\cite{KAT-15} and elliptical Ni nanodots in square arrays\cite{MAC-16}.

\begin{figure}[t]
	\includegraphics{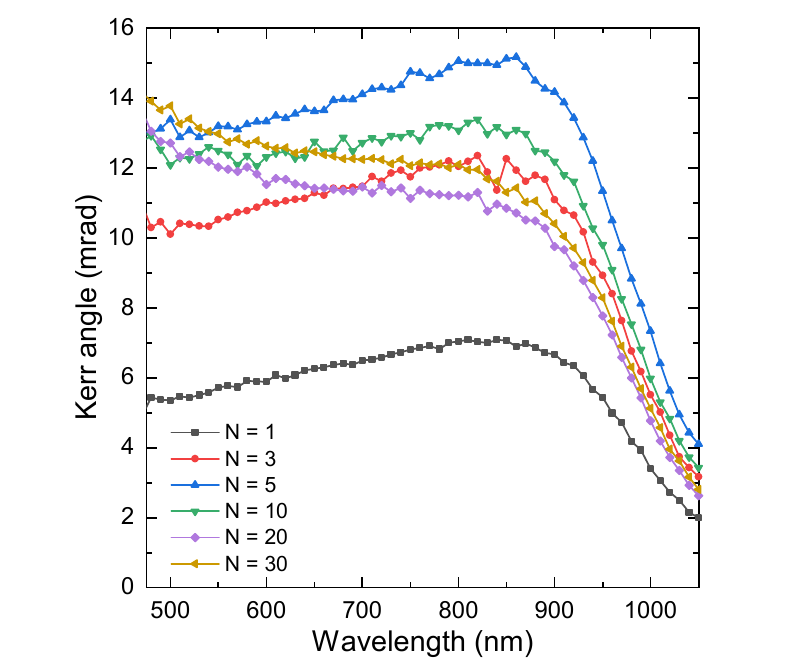}
	\caption{Polar MOKE spectra of continuous $[\textnormal{Co}/\textnormal{Pt}]_{N}$ multilayers as a function of $N$.}
	\label{Fig4}
\end{figure}

If we compare the optical reflectivity and MOKE spectra of our square $[\textnormal{Co}/\textnormal{Pt}]_{N}$ nanodot arrays (Fig. \ref{Fig3}), we indeed observe the expected resemblance. Excitation of the SLR mode produces a resonant enhancement of the magneto-optical Kerr angle. More intense SLR modes result in larger Kerr signals for most samples, as demonstrated by the dependence on nanodot diameter (Figs. \ref{Fig3}(a),(d)) and array period (Figs. \ref{Fig3}(b),(e)). However, the Kerr angle varies non-monotonically with the number of bilayer repetitions, with $N$ = 10 displaying the strongest MOKE signal (Fig. \ref{Fig3}(f)). To distinguish between plasmonic or magnetic effects that could be responsible for this, it is instructive to compare the MOKE spectra of the nanodot arrays to those of continuous $[\textnormal{Co}/\textnormal{Pt}]_{N}$ multilayers. Figure \ref{Fig4} shows that the MOKE signal decreases also with $N$ in continuous multilayers. From this we conclude that the upper Co layers in thick films exhibit a reduced magnetic moment, most likely caused by larger interface roughness. In the continuous multilayers, the Kerr angle already reduces for $N$ = 10. In the nanodot arrays, this intrinsic magnetic effect is initially compensated by the excitation of a more intense SLR mode for $N$ = 10 than $N$ = 5, producing a larger MOKE signal for $N$ = 10. To further quantify the resonant enhancement of the Kerr angle in $[\textnormal{Co}/\textnormal{Pt}]_{N}$ nanodot arrays, we compare the MOKE spectrum of $D$ = 200 nm, $P$ = 400 nm, and $N$ = 10 (black data in Fig. \ref{Fig3}(d)) to that of a continuous multilayer with $N$ = 10 (green data in Fig. \ref{Fig4}). In these two cases, the maximum Kerr angle is similar, 15.5 mrad versus 13 mrad. However, the packing density of the $[\textnormal{Co}/\textnormal{Pt}]_{10}$ nanodots is only 20\%. This suggest a SLR-induced resonant enhancement of the MOKE signal by a factor $\sim$6. Another way of assessing the effect of the SLR on the magneto-optical response of the nanodot arrays directly compares the Kerr angle measured on- and off-resonance. Using data of the same nanodot array we find that a small off-resonance signal of 1.6 mrad at 500 nm is resonantly enhanced by the SLR mode to 15.5 mrad at 800 nm (black curve in Fig. \ref{Fig3}(d)). 

\subsection{Optical and magneto-optical properties of $[\textnormal{Co}/\textnormal{Pt}]_{N}$ nanodot arrays on Si/Au/Si$\textnormal{O}_{2}$} 

While $[\textnormal{Co}/\textnormal{Pt}]_{N}$ nanodot arrays on glass enable resonant enhancements of their magneto-optical activity via the excitation of SLRs, the magneto-optical resonances are still relatively broad because of the strongly damped plasmonic component of SLR modes. Propagating surface plasmon polaritons (SPPs) excited at a noble metal/dielectric interface exhibit considerably lower damping. Free space photons cannot excite SPPs as their dispersion relation lies under the light line\cite{BAR-03}. To overcome this momentum mismatch, SPPs are often excited using a prism or nanostructure array that acts as grating coupler. In the experiments discussed in this section, we explore the optical and magneto-optical properties of $[\textnormal{Co}/\textnormal{Pt}]_{N}$ nanodot arrays on top of Au/Si$\textnormal{O}_{2}$ bilayers, wherein the Au and Si$\textnormal{O}_{2}$ layers are 150 nm and 20 nm thick, respectively. The rational behind this hybrid magnetoplasmonic structure is that the $[\textnormal{Co}/\textnormal{Pt}]_{N}$ array facilitates the excitation of low-loss SPPs at the Au/Si$\textnormal{O}_{2}$ interface. In turn, the slowly decaying near-fields of the SPP modes on the dielectric side of the Au/Si$\textnormal{O}_{2}$ interface induce a narrow-linewidth magneto-optical response on the $[\textnormal{Co}/\textnormal{Pt}]_{N}$ nanodots. Recently, this low-loss magnetoplasmonic excitation mechanism was demonstrated for the first time using Ni nanodisks\cite{FRE-19}.

The experiments on $[\textnormal{Co}/\textnormal{Pt}]_{N}$ nanodot arrays on top of Au/Si$\textnormal{O}_{2}$ bilayers are performed in reflection (the Au film blocks transmission). Absorption of light by the excitation of plasmon modes suppresses the optical reflectivity in this measurement geometry. The momentum mismatch between free space photons and SPPs is overcome by the extra momentum provided by the nanodot array\cite{EBB-98}

\begin{equation}
\textbf{k}_{SPP}=\textbf{k}_{0}+p\textbf{G}_{x}+q\textbf{G}_{y}.
\label{Eq5}
\end{equation}

\noindent Here, $\textbf{k}_{SPP}$ is wave vector of the SPP mode, $\textbf{k}_{0}=(\omega/c)\sin(\theta)$ is the wave vector of free space photons with angular frequency $\omega$ traveling at the speed of light $c$ and $\theta$ is the angle of incidence, $\textbf{G}_{x,y}=\frac{2\pi}{P_{x,y}}e_{x,y}$ are the reciprocal lattice vectors for array periods $P_{x,y}$ along the direction of unit vectors $e_{x,y}$, and $p,q$ indicate the order of diffraction along the $x$- and $y$-axis of the nanodot array. For a square array with period $P$ acting as grating coupler, the free space wavelength of the SPP mode corresponds to\cite{PAP-07}

\begin{equation}
\lambda'_{p,q}=\frac{P}{\sqrt{{p^2+q^2}}}\cdot\sqrt{\frac{\epsilon_\mathrm{d}\epsilon_\mathrm{m}}{\epsilon_\mathrm{d}+\epsilon_\mathrm{m}}},
\label{Eq6}
\end{equation}

\noindent where $\epsilon_\mathrm{m}$ and $\epsilon_\mathrm{d}$ are the dielectric constants of the metal film and the dielectric layer, respectively. In metal/dielectric bilayers with a periodic nanodot array on top, the SPP wavelength is greater than that of the DO. 

\begin{figure*}
	\includegraphics{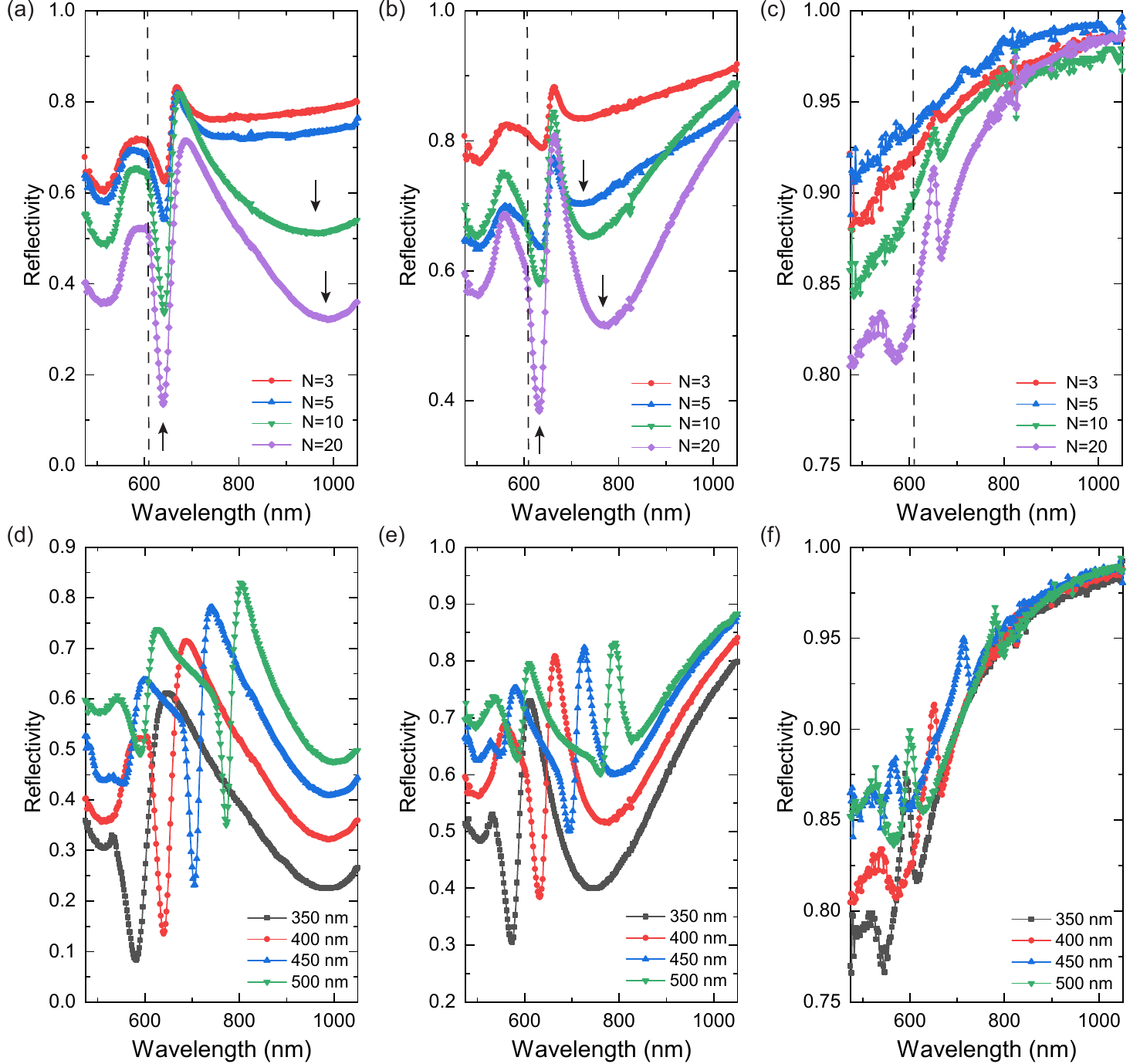}
	\caption{Optical reflectivity spectra of square arrays of $[\textnormal{Co}/\textnormal{Pt}]_{N}$ nanodots on a Au/Si$\textnormal{O}_{2}$ bilayer as a function of $N$ (a-c) and $P$ (d-f). Panels (a-c) show data for arrays with $P$ = 400 nm and $D$ = 200 nm (a), 150 nm (b), and 100 nm (c). The vertical dashed lines denote the position of the (0,$\pm$1) DOs, the upward pointing arrows mark the SPP mode, and the downward pointing arrows indicate the SLR. Panels (d-f) depict data for arrays with $N$ = 20 nm and $D$ = 200 nm (d), 150 nm (e), and 100 nm (f).}
	\label{Fig5}
\end{figure*}

Figures \ref{Fig5}(a)-(c) show optical reflectivity spectra of $[\textnormal{Co}/\textnormal{Pt}]_{N}$ nanodot arrays on a Au/Si$\textnormal{O}_{2}$ bilayer. The array period is 400 nm in panels (a)-(c) and $D$ = 200 nm (a), 150 nm (b), and 100 nm (c). Two clear reflectivity minima are measured. The first narrow resonance marked by an upward pointing arrow occurs just above the wavelength of the (0,$\pm$1) DOs of the array (vertical dashed line in Figs. \ref{Fig5}(a),(b)). This resonance corresponds to a SPP mode. Based on Eq. \ref{Eq6} and the dielectric constants of Au and Si$\textnormal{O}_{2}$ determined by ellipsometry, we estimate an SPP excitation wavelength of 638 nm for $P$ = 400 nm, in excellent agreement with the experimental data of Figs. \ref{Fig5}(a),(b). The second reflectivity minimum occurring at larger wavelength is considerably broader (labeled by a downward pointing arrow). This resonance corresponds to the SLR mode in the $[\textnormal{Co}/\textnormal{Pt}]_{N}$ nanodot array. Compared to the same array on glass, the SLR mode on Au/Si$\textnormal{O}_{2}$ is red-shifted (compare Figs. \ref{Fig5}(a),(b) and Figs. \ref{Fig2}(a),(b)). This effect is explained by the formation of image dipoles in the Au film when the Si$\textnormal{O}_{2}$ layer is thin, as discussed previously for noble metal plasmonic systems\cite{HOL-84,NOR-04}. Energy absorption by the SPP and SLR modes increases with $N$ and $D$. Figures \ref{Fig5}(d)-(f) summarize tuning of the optical reflectivity spectra by the array period. In these measurements, $N$ = 20. Since the SPP and SLR modes both depend on the spectral position of the DOs, the two reflectivity minima red-shift with increasing $P$.

\begin{figure}
	\includegraphics{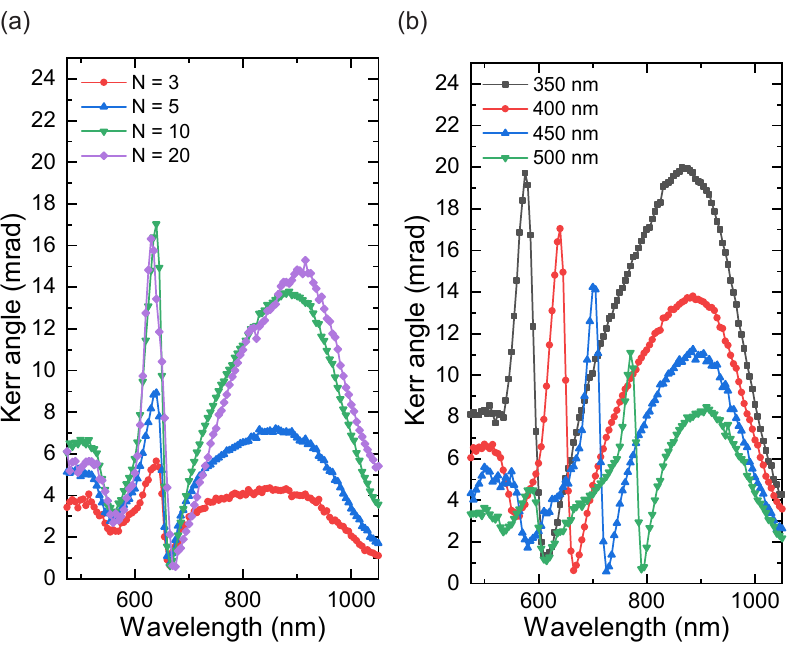}
	\caption{Polar MOKE spectra of square arrays of $[\textnormal{Co}/\textnormal{Pt}]_{N}$ nanodots on a Au/Si$\textnormal{O}_{2}$ bilayer as a function of $N$ (a) and $P$ (b). In (a) $D$ = 200 nm and $P$ = 400 nm. In (b) $D$ = 200 nm and $N$ = 10.}
\label{Fig6}
\end{figure}

We now focus our attention to the magneto-optical response of the $[\textnormal{Co}/\textnormal{Pt}]_{N}$ nanodot arrays on Si/Au/Si$\textnormal{O}_{2}$. Figure \ref{Fig6} shows the Kerr angle dependence on the number of Co/Pt bilayer repetitions and the array period. Obviously, the SPP and SLR modes both produce a strong resonant enhancement of the MOKE signal. The SLR-induced Kerr angle and its dependence on $N$ and $P$ is similar to that observed in $[\textnormal{Co}/\textnormal{Pt}]_{N}$ nanodot arrays on glass, except for a red-shift of the resonances caused by the formation of image dipoles in the Au film. Section B describes how SLR modes produce a strong MOKE signal in magnetic nanodot arrays. Here, we focus on the magneto-optical activity of the SPP mode. The linewidth of the SPP-induced MOKE resonance is small, with the full width at half maximum (FWHM) ranging from 54 nm ($D$ = 200 nm, $P$ = 350 nm) to 18 nm ($D$ = 200 nm, $P$ = 500), which is similar to the SPP resonances linewidth in the optical reflectivity spectra of Fig. \ref{Fig5}. While the SPP mode is excited at the Au/Si$\textnormal{O}_{2}$ interface, its near-field only decays slowly within the dielectric film. Typically, the decay length of the SPP electric field is about half the wavelength of light involved\cite{BAR-03}. Because the Si$\textnormal{O}_{2}$ layer in our magnetoplasmonic structures is much thinner than this decay length, the $[\textnormal{Co}/\textnormal{Pt}]_{N}$ nanodots patterned on top of Si$\textnormal{O}_{2}$ are driven into resonance. Consequently, the SPP mode induces an electric dipole ($p_\mathrm{x}$) in the magnetic nanodots parallel to the incident electric field ($E_\mathrm{x}$). Via spin-orbit coupling in $[\textnormal{Co}/\textnormal{Pt}]_{N}$ this again produces an orthogonal electric dipole ($p_\mathrm{y}$), rendering a magneto-optical signal at the SPP wavelength. Since the SPP resonances forces the free electrons of the $[\textnormal{Co}/\textnormal{Pt}]_{N}$ nanodots into oscillation, the induced magneto-optical response is not strongly affected by damping in the nanodots. This point, illustrating a powerful loss mitigation strategy for metallic magnetoplasmonics, was recently substantiated by the demonstration of similar SPP linewidths for Ni and Au nanodot arrays on Au/Si$\textnormal{O}_{2}$ bilayers\cite{FRE-19}.   

The magnitude of the Kerr angle at the SPP wavelength increases initially with the number of Co/Pt bilayer repetitions before it saturates at $N$ = 10 - 20 (Fig. \ref{Fig6}(a)). This effect, which is similar to that observed for the SLR mode, again correlates with a reduction of the magnetic moment in the upper Co layers of thick multilayers (Fig. \ref{Fig4}). The maximum Kerr angle measured at the SPP wavelength is comparable to that produced by the SLR ($\sim$20 mrad for $N$ = 10, $D$ = 200 nm, and $P$ = 350 nm). Variation of the array period offers an attractive means of tuning the wavelength of the narrow SPP-induced MOKE resonance. The spectral position of the Kerr angle maximum is given by Eq. \ref{Eq6}, which can be used as a design tool in magnetoplasmonic applications. The decrease of the Kerr angle with array period (Fig. \ref{Fig6}(b)) is caused predominantly by a reduction of the nanodot packing density from 26\% for $P$ = 350 nm to 13\% for $P$ = 500 nm.                   
 
\section{Summary}
\label{Sec:Conclusions}

In this article, we report on the magnetoplasmonic properties of perpendicularly magnetized $[\textnormal{Co}/\textnormal{Pt}]_{N}$ nanodot arrays on glass and Si/Au/Si$\textnormal{O}_{2}$. On glass, the optical and magneto-optical responses are dominated by the excitation of a collective SLR mode, arising from diffractive coupling between single-particle LSPRs. On Si/Au/Si$\textnormal{O}_{2}$, a red-shifted SLR mode is also measured. In addition, a spectrally more narrow resonance appears in optical reflectivity and MOKE spectra. This feature is induced by the excitation of a SPP mode at the Au/Si$\textnormal{O}_{2}$ interface. Because the $[\textnormal{Co}/\textnormal{Pt}]_{N}$ nanodots are placed within the SPP near-field, plasmon resonances are excited in the nanodots at the SPP wavelength. In both cases, whether plasmon resonances in the nanodots are directly excited by the incident electric field or via a SPP, spin-orbit coupling produces a second plasmon mode in the orthogonal direction. This effect causes linear polarized light to undergo a rotation and to become elliptical upon reflection from the $[\textnormal{Co}/\textnormal{Pt}]_{N}$ nanodot samples. 

Compared to continuous $[\textnormal{Co}/\textnormal{Pt}]_{N}$ films or off-resonance measurement conditions, optical near-field enhancements at the SPP and SLR wavelength can increase local magneto-optical effects by up to one order of magnitude. Plasmon-enhanced MOKE signals require a $[\textnormal{Co}/\textnormal{Pt}]_{N}$ nanodot diameter of more than 100 nm and 10 - 20 Co/Pt bilayer repetitions maximize the Kerr angle. Because the SLR mode depends on single-particle LSPRs, its spectral position is tuned by variation of the multilayer thickness and the nanodot diameter. Both the SLR and SPP modes depend on the DOs of the array and, consequently, their spectral position depends on the array period. 

Plasmon-enhanced magneto-optical effects in $[\textnormal{Co}/\textnormal{Pt}]_{N}$ nanodots and the flexibility to tailor strong light-matter interactions in this PMA material system may be utilized in AOS experiments or as an efficient interface linking photonic and spintronic devices\cite{LAL-19}. Our findings could also be exploited in plasmonic lasing. Recently, lasing was demonstrated for the first time in a magnetic system\cite{POUR-19}. In this study, the SLR mode of Ni nanodot arrays on glass acted as a feedback mechanism for lasing from an organic gain medium. We expect that the much narrower SPP resonances in $[\textnormal{Co}/\textnormal{Pt}]_{N}$ nanodot arrays on Au/Si$\textnormal{O}_{2}$ would provide more efficient feedback, with the potential of realizing magnetic field control of plasmonic lasing.     
                  
\section{Acknowledgements} 
This work was supported by the Academy of Finland (Grant Nos. 316857, 295269, and 306978). F.F-F. acknowledges financial support from the Finnish Academy of Science and Letters (Vilho, Yrjö and Kalle Väisälän Fund). Lithography was performed at the Micronova Nanofabrication Centre, supported by Aalto University. 


%

\end{document}